\documentclass[sigconf]{acmart}

\usepackage{booktabs}
\usepackage{subcaption}
\usepackage{listings}
\usepackage{threeparttable}
\usepackage{multirow}
\usepackage{mwe,tikz}
\usepackage{siunitx}
\usepackage{xcolor,colortbl}
\usepackage{xparse}
\usepackage{enumitem}

\newcommand{\titlename}{$f$\kern-0.18em\emph{unc}\kern-0.05em X}
\newcommand{\name}{$f$\kern-0.18em\emph{unc}\kern-0.05em X}

\newcommand{\dlhub}{DLHub}
\newcommand{\xtract}{Xtract}
\newcommand{\parsl}{Parsl}

\newsavebox{\fminipagebox}
\NewDocumentEnvironment{fminipage}{m O{\fboxsep}}
 {\par\kern#2\noindent\begin{lrbox}{\fminipagebox}
  \begin{minipage}{#1}\ignorespaces}
 {\end{minipage}\end{lrbox}%
  \makebox[#1]{%
    \kern\dimexpr-\fboxsep-\fboxrule\relax
    \fbox{\usebox{\fminipagebox}}%
    \kern\dimexpr-\fboxsep-\fboxrule\relax
  }\par\kern#2
 }

 \definecolor{deepred}{rgb}{0.6,0,0}
 \definecolor{deepgreen}{rgb}{0,0.5,0}

\lstdefinestyle{PythonStyle}
{
        language=Python,
        frame=tb,
        basicstyle=\ttfamily\small,
        upquote=true,
        numbers = none,
        escapechar=`,
        float,
        moredelim=[il][]{--latexlabel},
        otherkeywords={self},             
        commentstyle=\color{blue},
        keywordstyle=\bfseries\color{black},
        emph={MyClass,__init__,Config,HighThroughputExecutor,SlurmProvider,LocalChannel}, 
        emphstyle=\bfseries\color{deepred},    
        showstringspaces=false            %
}

\lstdefinestyle{PythonStyleInLine}
{
        language=Python,
        basicstyle=\small\ttfamily,
        upquote=true,
        numbers = none,
        numberstyle=\footnotesize,
        escapechar=`,
        moredelim=[il][]{--latexlabel},
        otherkeywords={self},             
        commentstyle=\color{blue},
        keywordstyle=\bfseries\color{red},
        emph={MyClass,__init__,@python_app,@bash_app},          
        emphstyle=\bfseries\color{deepred},    
        showstringspaces=false,            %
        frame=bt
}

\newif\iffinal

\iffinal
  \newcommand{\ian}[1]{}
  \newcommand{\ryan}[1]{}
  \newcommand{\kyle}[1]{}
  \newcommand{\zhuozhao}[1]{}
  \newcommand{\yadu}[1]{}
  \newcommand{\tyler}[1]{}
  \newcommand{\ben}[1]{}
  \newcommand{\anna}[1]{}
  \newcommand{\TODO}[1]{}
\else
  \newcommand{\ian}[1]{{\textcolor{red}{ Ian: #1 }}}
  \newcommand{\TODO}[1]{{\textcolor{red}{ TODO: #1 }}}
  \newcommand{\ryan}[1]{{\textcolor{magenta}{ Ryan: #1 }}}
  \newcommand{\kyle}[1]{{\textcolor{purple}{ Kyle: #1 }}}
  \newcommand{\zhuozhao}[1]{{\textcolor{cyan}{ Zhuozhao: #1 }}}
  \definecolor{darkgreen}{rgb}{0,0.5,0}
  \newcommand{\tyler}[1]{{\textcolor{green}{ Tyler: #1 }}}
  \newcommand{\yadu}[1]{{\textcolor{orange}{ Yadu: #1 }}}
  \definecolor{pink}{rgb}{1.0,0,0.5}
  \newcommand{\ben}[1]{{\textcolor{deepgreen}{ Ben: #1 }}}
  \newcommand{\anna}[1]{{\textcolor{deepred}{ Anna: #1 }}}

\fi

\newif\ifchange

\ifchange
  \newcommand{\change}[1]{ #1 }
\else
  \newcommand{\change}[1]{{\textcolor{black}{ #1 }}}
\fi

\newcommand\mycode[1]{{\texttt{\small #1}}}

\author{Ryan Chard}
\affiliation{Argonne National Laboratory}

\author{Yadu Babuji}
\affiliation{University of Chicago}

\author{Zhuozhao Li}
\affiliation{University of Chicago}

\author{Tyler Skluzacek}
\affiliation{University of Chicago}

\author{Anna Woodard}
\affiliation{University of Chicago}

\author{Ben Blaiszik}
\affiliation{University of Chicago}

\author{Ian Foster}
\affiliation{Argonne National Laboratory and University of Chicago}

\author{Kyle Chard}
\affiliation{University of Chicago and Argonne National Laboratory}

\renewcommand{\shortauthors}{R. Chard et al.}

\setcopyright{none}
\renewcommand\footnotetextcopyrightpermission[1]{}
\begin{document}
\fancyhead{}

\title[]{\titlename{}: A Federated Function Serving Fabric for Science}


\begin{abstract}
Exploding data volumes and velocities, new computational methods and platforms, 
and ubiquitous connectivity demand new approaches to computation in the sciences. 
These new approaches must enable computation to be mobile, so that, for example, 
it can occur near data, be triggered by events (e.g., arrival of 
new data), be offloaded to specialized accelerators, or run remotely
where resources are available. 
They also require new design approaches in which monolithic applications 
can be decomposed into smaller components, that may in turn
be executed separately and on the most suitable resources. 
To address these needs we present \name{}---a distributed  
function as a service (FaaS) platform that enables flexible, 
scalable, and high performance remote function execution. 
\name{}'s endpoint software can transform existing clouds, clusters, 
and supercomputers into function serving systems, while 
\name{}'s cloud-hosted service provides transparent, secure, 
and reliable function execution across a federated ecosystem of endpoints. 
We motivate the need for \name{} with several scientific case studies, present 
our prototype design and implementation, 
show optimizations that deliver throughput in excess of 1 million functions per second,
and demonstrate, via experiments on two supercomputers,
that \name{} can scale to more than more than \num{130000} concurrent workers.

\end{abstract}

\maketitle

\section{Introduction}

The idea that one should be able to compute wherever makes the most sense---wherever a suitable computer
is available, software is installed, or data are located, for example---is far from new: indeed, it predates the Internet~\cite{fano1965mac,parkhill1966challenge}, and motivated initiatives such as grid~\cite{Foster2001} and peer-to-peer computing~\cite{milojicic2002peer}. 
But in practice remote computing has long been complex and expensive, due to, for example, 
slow and unreliable network communications, security challenges, and heterogeneous computer architectures.

Now, however, with ubiquitous high-speed communications, universal trust fabrics, and containerization,
computation can occur essentially anywhere. 
Commercial cloud services have embraced this new reality~\cite{varhese19cloud},
in particular via their
function as a service (FaaS)~\cite{baldini2017serverless,fox2017status} offerings 
that make invoking remote functions trivial.
Thus one simply writes \mycode{client.invoke(FunctionName="F", Payload=D)} to invoke a remote 
function \mycode{F(D)}.
The FaaS model allows monolithic applications to be transformed into ones that use
event-based triggers to dispatch tasks to remote cloud providers. 

There is growing awareness of the benefits of decomposing monolithic 
science applications into functions that can be more
efficiently executed on
remote computers~\cite{foster2017cloud,spillner2017faaster,malawski2016towards,fox2017conceptualizing,kiar2019serverless}. 
As we move towards this reality, it becomes easy for scientists to 
place computations wherever it makes the most sense, and to then
move those computations between resources. 
For example, physicists at FermiLab report that a data analysis task 
that takes two seconds on a CPU can be dispatched to an FPGA device on 
the Amazon Web Services (AWS) cloud, where it takes 30 ms to execute, for a total of 50 ms 
once a round-trip latency of 20 ms to Virginia is included: 
a speedup of 40$\times$~\cite{duarte2018fast}.
Such examples can be found in many scientific domains; however, 
until now, there has been no universal and easy-to-use way to remotely execute
and move functions between resources.

Unfortunately, existing FaaS systems are not designed to be deployed
on heterogeneous research cyberinfrastructure (CI) nor are they
designed to federate resources. Further, existing CI is 
not designed to support granular and sporadic function execution. 
For example, existing CI typically expose batch scheduling interfaces, 
use inflexible authentication and authorization models, and have unpredictable 
scheduling delays for provisioning resources, to name just a few challenges. 
We are therefore motivated to overcome these challenges by
adapting the FaaS model to research CI with the aim to 
enable reliable computation of granular tasks (i.e., at the level 
of programming functions) at scale across a diverse 
range of existing CI, including clouds, clusters, and supercomputers. 

To explore this approach we have developed a distributed, scalable, 
and high-performance function execution platform, \name{}, that adapts the powerful and flexible FaaS model to support science workloads across federated research CI, a model that is not achievable with existing FaaS platforms.
\name{} is a cloud-hosted software as a service (SaaS) system
that allows researchers to register Python functions 
and then invoke those functions on supplied inputs on remote CI.
\name{} manages the reliable and secure execution of functions on remote CI, 
provisioning resources, staging function code and inputs, managing safe and secure
execution sandboxes using containers, monitoring execution, and returning outputs to users. 
Functions can execute on any compute resource where 
\name{}'s endpoint software, a \name{} agent, is installed
and that a requesting user is authorized to access. 
\name{} agents can turn \emph{any} existing resource (e.g., laptop, cloud, cluster, supercomputer, 
or container orchestration cluster) into a FaaS endpoint.

The primary novelty of \name{} lies in how it
combines the extreme convenience of FaaS
with support for the specialized and distributed research ecosystem. 
On-demand remote computing, of which FaaS is an implementation,
has motivated initiatives such as grid~\cite{Foster2001}, Condor~\cite{thain05condor},
peer-to-peer computing~\cite{milojicic2002peer}, and Remote Procedure Call (RPC) implementations. 
The cloud-based FaaS that dominates in industry innovates by enabling on-demand function execution on cloud datacenters. 
Open source FaaS systems enable function execution on a single computer or
container orchestration system~\cite{baldini2016cloud,Fn,Kubeless,stubbs2017containers}.

\name{} is the first federated FaaS system 
that enables execution of functions across heterogeneous, distributed resources. 
Building on a hybrid cloud model that combines
user-managed endpoints and a reliable cloud management service, it
implements a multi-layered and reliable communication model 
to overcome the unreliability of distributed endpoints; 
supports heterogeneous resources with various cloud and batch scheduler interfaces, 
container technologies, and unpredictable provisioning delays;
integrates with the research identity and data management ecosystem 
providing access via standard web authorization protocols;
and includes performance optimizations focused on addressing the unique 
challenges associated with this federated FaaS model. 
\name{} thus serves as a foundational research platform on which a range of 
new applications can be developed and research opportunities explored, from 
multi-level function scheduling and hybrid cloud-edge computing, to 
scalable data management and integration of accelerators.

The contributions of our work are as follows: 
\begin{itemize}
    \item The distributed and federated \name{} platform that can: 
          be deployed on research CI, dynamically provision and manage
					resources, use various container technologies, and facilitate 
					secure, scalable, and distributed function execution.
    \item Design and evaluation of performance enhancements for function
          serving on research CI, including memoization, function warming, 
					batching, and prefetching.
    \item Experimental studies showing that \name{} delivers execution 
					latencies comparable to those of commercial FaaS platforms 
          and scales to 1M+ functions across 130K active workers on supercomputers.
    \item Discussion of our experiences applying \name{} to real-world use
			cases and exploration of the advantages and disadvantages of FaaS in science.
\end{itemize}

The rest of this paper is as follows. 
\S\ref{sec:requirements} describes requirements of FaaS in science.
\S\ref{sec:funcx} presents a conceptual model of \name{}.
\S\ref{sec:arch} describes the \name{} system architecture. 
\S\ref{sec:evaluation} evaluates \name{} performance. 
\S\ref{sec:discussion} reviews \name{}'s use in scientific case studies. 
\S\ref{sec:survey} discusses related work.
Finally, \S\ref{sec:conclusion} summarizes our contributions.

\section{Requirements}
\label{sec:requirements}

\begin{figure}[ht]
  \includegraphics[width=1\columnwidth,trim=0.08in 0.1in 0.08in 0.1in,clip]{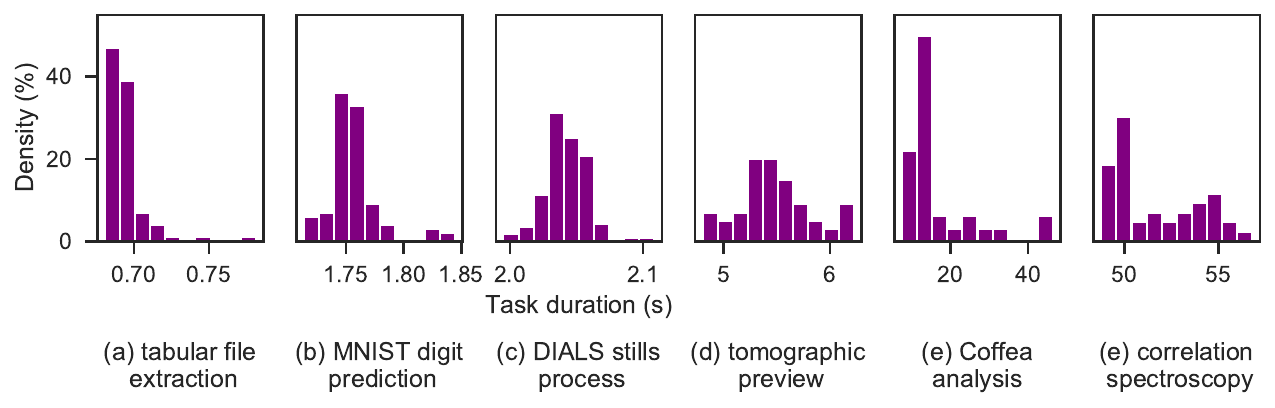}
    \vspace{-0.15in}
  \caption{Distribution of latencies for 100 function calls, for each of the six case studies described in the text.
\label{fig:use-cases}
  \vspace{-0.1in}
}
\end{figure}

Our work is guided by the unique requirements of FaaS in science. 
To illustrate these requirements we present six representative case studies:
scalable metadata extraction, machine learning inference as a service, synchrotron serial crystallography, neuroscience, correlation spectroscopy, and high energy physics.  
\figurename~\ref{fig:use-cases} shows function execution time distributions for each case study. 
These short duration tasks exemplify opportunities for FaaS in science. 
We summarize by highlighting requirements for FaaS in science.

\textbf{Metadata extraction:}
The effects of high-velocity data expansion are making it increasingly
difficult to organize, discover, and manage data. 
Edge file systems and data
repositories now store petabytes of data
which are created and modified at an alarming rate~\cite{paul17monitoring}.
\xtract{}~\cite{skluzacek19serverless} 
is a distributed metadata extraction system
that applies a set of
general and specialized metadata extractors, such as those for identifying topics in text, computing aggregate values from tables, and recognizing locations in maps. 
To reduce data transfer costs, \xtract{} executes extractors ``near'' to
data, by pushing extraction tasks to the edge. Extractors are implemented
as Python functions, with various dependencies, and each extractor typically executes for between 3 milliseconds and 15 seconds.

\textbf{Machine learning inference:}
\dlhub{}~\cite{chard19dlhub} 
is a service that supports ML model publication 
and on-demand inference. 
Users deposit ML models, implemented as functions with a set of
dependencies, in the \dlhub{} catalog
by uploading the raw model (e.g., PyTorch, TensorFlow) and model 
state (e.g., training data, hyperparameters).
\dlhub{} uses this information to dynamically create a container for the model using repo2docker~\cite{repo2docker} that contains all model dependencies and necessary model
state.
Other users may then invoke the model through \dlhub{} on arbitrary input
data. \dlhub{} currently includes more than one hundred published models, 
many of which have requirements for specific ML toolkits and execute
more efficiently on GPUs and accelerators. 
\figurename~\ref{fig:use-cases} shows the execution time when invoking the 
MNIST digit identification model. Other \dlhub{} models execute for between seconds and
several minutes. 

\textbf{Synchrotron Serial Crystallography (SSX)} 
is an emerging method for imaging small crystal samples 1--2 orders of magnitude faster than other methods. 
To keep pace with the increased data production, SSX researchers require automated 
approaches that can process the resulting data with great rapidity:
for example, to count the \emph{bright spots} in an image 
(``stills processing'') within seconds, both for quality control 
and as a first step in structure determination. The DIALS~\cite{waterman2013dials} crystallography processing tools are implemented
as Python functions that execute for 
1--2 seconds per sample. 
Analyzing large datasets requires HPC resources to derive crystal structures
in a timely manner.

\textbf{Quantitative Neurocartography} and connectomics map the 
neurological connections in the brain---a compute- and data-intensive process 
that requires processing \textasciitilde 20GB every minute during experiments.
Researchers apply automated workflows 
to perform quality control on raw images (to validate that the instrument
and sample are correctly configured), apply ML models
to detect image centers for subsequent reconstruction, and generate
preview images to guide positioning. Each of these steps is implemented
as a function that can be executed sequentially with some data exchange
between steps. However, given the significant data sizes, researchers typically
rely on HPC resources and are subject to scheduling delays.

\textbf{Real-time data analysis in High Energy Physics (HEP)}:
The traditional HEP analysis model uses successive processing steps to reduce the initial dataset (typically, 100s of PB) to a size that permits real-time analysis. This iterative approach requires significant computation time and storage of large intermediate datasets, and may take weeks or months to complete. Low-latency, query-based analysis strategies~\cite{pivarski2017} are being developed to enable real-time analysis
using native operations on hierarchically nested, columnar data. 
Such queries are well-suited to FaaS. 
To enable interactive analysis, for example as a physicist engages in real-time analysis of several billion particle collision events, successive compiled functions, each running for seconds, need to be dispatched to the data. Analysis needs require sporadic (and primarily remote) invocation, and compute needs increase as new data are collected.

\textbf{X-ray Photon Correlation Spectroscopy (XPCS)} 
is an experimental technique used 
to study the dynamics in materials at nanoscale by identifying correlations in 
time series of area detector images. 
This process involves analyzing the pixel-by-pixel correlations 
for different time intervals. 
The detector acquires megapixel frames at 60 Hz (\textasciitilde 120 MB/sec). 
Computing correlations at these data rates is a challenge that requires HPC resources but also rapid response time. Image processing functions, such as XPCS-eigen's corr function,
execute for approximately 50 seconds, and images can be processed in parallel.

\textbf{Requirements for FaaS in science:}
The case studies illuminate benefits of 
FaaS approaches (e.g., decomposition, abstraction, 
flexibility, scalability, reliability), but also
elucidate requirements unmet by existing FaaS solutions: 

\begin{itemize} 
    \item \textbf{Specialized compute:} functions may require HPC-scale and/or specialized and heterogeneous resources (e.g., GPUs).   
    \item \textbf{Distribution:} functions may need to execute near to data and/or on a specialized computer. 
  	\item \textbf{Dependencies:} functions often require specific libraries and user-specified dependencies.
	\item \textbf{Data:} functions analyze both small and large data, stored in various locations and formats, and accessible via different methods (e.g., Globus~\cite{chard14efficient}).
    \item \textbf{Authentication:} institutional identities and specialized security models are used to access data and compute resources.
    \item \textbf{State:} functions may be connected and share state (e.g., files or database connections) to decrease overheads. 
    \item \textbf{Latency:} functions may be used in online (e.g., experiment steering) and interactive environments (e.g., Jupyter notebooks) that require rapid response. 
    \item \textbf{Research CI:} resources offer batch scheduler interfaces (with long delays, periodic downtimes, proprietary interfaces) and specialized container technology (e.g., Singularity, Shifter) that make it challenging to provide common execution interfaces, elasticity, and fault tolerance. 
    \item \textbf{Billing:} research CI use allocation-based usage models.
\end{itemize}

\section{Conceptual Model}\label{sec:funcx}

We first describe the conceptual model behind \name{}
to provide context to the implementation architecture. 
\name{} allows users to register and then execute \emph{functions} 
on arbitrary \emph{endpoints}.
All user interactions with \name{} are performed via a 
REST API implemented by a cloud-hosted \name{} service.

\begin{lstlisting}[style=PythonStyle, caption={Using \name{} SDK to register and invoke a function. \label{lst:sdk}}]
def automo_preview(fname, start, end, step):
  import numpy, tomopy
  from automo.util import read_adaptive, save_png
  proj, flat, dark, _ = read_adaptive(
      fname, proj=(start, end, step))
  proj_norm = tomopy.normalize(proj, flat, dark)
  flat = flat.astype('float16')
  save_png(flat.mean(axis=0), fname='prev.png')
  return 'prev.png'

fc = FuncXClient()
func_id= fc.register_function(automo_preview)
endpoint_id = '863d-...-d820d'

task_id = fc.run(func_id, endpoint_id, 
        fname='test.h5', start=0, end=10, step=1)
res = fc.get_result(task_id)
\end{lstlisting} 

\textbf{Functions:}
\name{} is designed to execute \emph{functions}---snippets of Python code
that perform some activity. A \name{} function explicitly defines
a Python function and input signature. 
The function body must specify all imported modules.
Listing~\ref{lst:sdk} shows the registration of a function for creating a tomographic 
preview image from raw tomographic data contained in an
HDF5 input file. The function's input specifies the file and 
parameters to identify and read a projection. 
It uses the Automo~\cite{Automo} Python package to read the data, 
normalize the projection, and then save the preview image. 
The function returns the name of the saved preview image.

\textbf{Function registration:}
A function must be registered with the \name{} service before it can be executed. 
Registration is performed via a JSON POST request to the REST API.
The request includes: a name and the serialized function body. 
Users may also specify users, or groups of users, who may invoke the function. 
Optionally, the user may
specify a container image to be used. Containers 
allow users to construct environments with appropriate dependencies
(system packages and Python libraries) required to execute the function.
\name{} assigns a universally
unique identifier (UUID) for management and invocation. Users may update
functions they own. 

\textbf{Endpoints:}
A \name{} endpoint is a logical entity that represents a compute resource.
The corresponding \name{} agent allows the \name{} service to dispatch 
functions to that resource for execution. The agent handles 
authentication and authorization, provisioning of nodes on the
compute resource, and monitoring and management. Administrators
or users can deploy a \name{} agent and register an endpoint for themselves and/or others,
providing descriptive (e.g., name, description) metadata. 
Each endpoint is assigned a unique identifier for subsequent use.

\textbf{Function execution:}
Authorized users may invoke a registered function on a selected endpoint. 
To do so, they issue a request via the \name{} service which identifies 
the function and endpoint to be used as well as inputs
to be passed to the function.   
Functions are executed asynchronously: each invocation returns an
identifier via which progress may be monitored and results retrieved. 
We refer to an invocation of a function as a ``task.''

\textbf{\name{} service:}
Users interact with \name{} via a cloud-hosted service which
exposes a REST API for registering functions and 
endpoints, and for executing functions, monitoring their execution, and retrieving results. 
The service is paired with accessible endpoints via the endpoint
registration process.

\textbf{User interface:}
\name{} provides a Python SDK that wraps the REST API. 
Listing~\ref{lst:sdk} shows an example of how the SDK can be used
to register and invoke a function on a specific endpoint.
The example first constructs a \emph{client} and registers the preview function. 
It then invokes the registered function using the \texttt{run}
command and passes the unique function identifier, 
the endpoint id on which to execute the function, and inputs
(in this case \texttt{fname}, \texttt{start}, \texttt{end}, and \texttt{step}). 
Finally, the example shows that the asynchronous results can be retrieved
using \texttt{get\_result}.

\section{Architecture and Implementation}\label{sec:arch}
\begin{figure}[h]
  \includegraphics[width=\columnwidth,trim=0.2in 0.1in 1.6in 0in,clip]{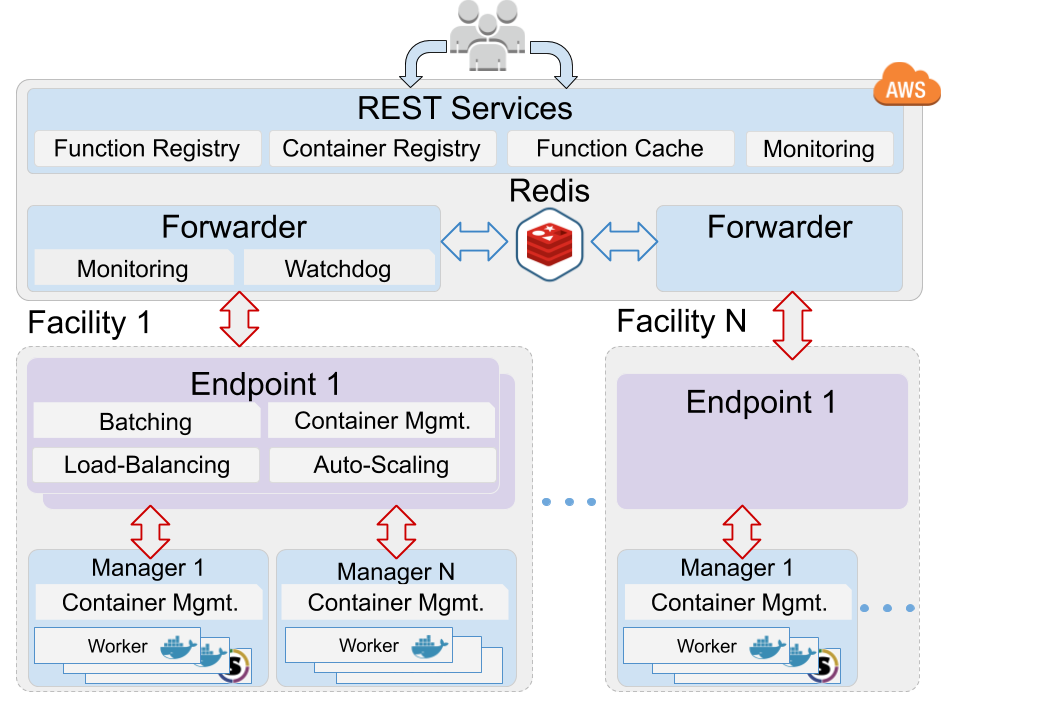}
  \vspace{-0.1in}
  \caption{\name{} architecture showing the \name{} service (top) consisting of a REST interface, Redis store, and Forwarders. \name{} endpoints (bottom) provision resources and coordinate the execution of functions. 
  }
\label{fig:arch}
\vspace{-0.1in}
\end{figure}

The \name{} system combines a cloud-hosted management service with software agents
deployed on remote resources: see  \figurename{~\ref{fig:arch}}.

\subsection{The \name{} Service}
The \name{} service maintains a registry of \name{} endpoints,
functions, and users in a persistent AWS Relational Database Service (RDS) database. 
To facilitate rapid function dispatch, \name{} stores serialized function codes 
and tasks (including inputs and task metadata) in an AWS ElastiCache Redis hashset.
The service also manages a Redis queue for each endpoint that stores task ids for
tasks to be dispatched to that endpoint.
The service provides a REST API to register and manage endpoints, 
register functions, execute and monitor functions, and retrieve the output from tasks. 
The \name{} service is secured using Globus Auth~\cite{GlobusAuth} 
which allows users, programs and applications, and \name{} endpoints
to securely make API calls. 
When an endpoint registers with the \name{} service
a unique \emph{forwarder} process is created for each 
endpoint. 
Endpoints establish ZeroMQ connections with their 
forwarder to receive tasks, 
return results, and perform heartbeats.

\name{} implements a hierarchical task queuing architecture consisting 
of queues at the \name{} service, endpoint, and worker. These queues support 
reliable fire-and-forget function execution that is resilient to failure 
and intermittent endpoint connectivity. 
At the first level, 
each registered endpoint is allocated a unique Redis \emph{task queue} 
and \emph{result queue} that reliably stores and tracks tasks. 

\begin{figure}[h]
  \includegraphics[width=\columnwidth]{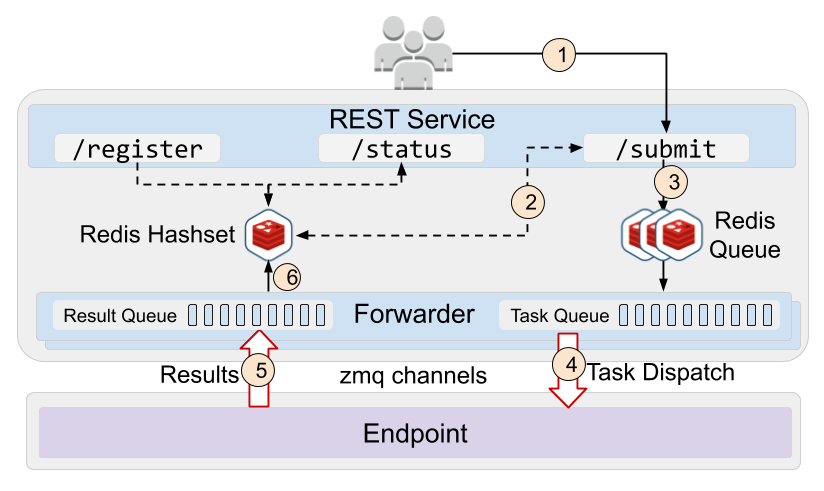}
  \vspace{-0.3in}
  \caption{\name{} task execution path. A task transmitted to \name{} (1) is stored in Redis (2), queued for execution (3), and dispatched via a Forwarder to an endpoint (4); results are returned (5) then stored in Redis for users to retrieve (6).
  }
\label{fig:taskpath}
\vspace{-0.1in}
\end{figure}

\figurename~\ref{fig:taskpath} shows the \name{} task lifecycle.
At function submission the \name{} service routes the task to the specified endpoint's task queue.
The forwarder dispatches tasks to the agent only when an agent is connected. The forwarder uses
heartbeats to detect if an agent is disconnected and then returns outstanding tasks back into the 
task queue. When the agent reconnects the tasks are forward to that agent.
This architecture ensures that \name{} agents receive tasks with at least once semantics. 
\name{} agents internally queue tasks at both the manager and worker.
These queues ensure that tasks are not lost once they
have been delivered to the endpoint. 
Similarly, results are returned to the \name{} service and stored in the endpoint's
result queue until they are retrieved by the user.

\change{\name{} relies on AWS hosted databases, caches, and Web serving infrastructure 
to reduce operational overhead,
elastically scale resources, and provide high availability. 
While these services provide significant benefits to \name{}, they have associated costs. 
To minimize these costs we apply several techniques, such as
using small cloud instances with responsive 
scaling to minimize the steady state
cost 
and restricting the size of input and output data passed through the 
\name{} service to reduce storage costs (e.g., in Redis store). 
For larger data sizes we use out-of-band transfer mechanisms
such as Globus~\cite{chard14efficient}. 
Further, we periodically purge results from the Redis store 
once they have been retrieved by the client. 
}

\subsection{Function Containers} 
\name{} uses containers to package function code and dependencies that are to be deployed
on a compute resource. 
Our review of container technologies, including Docker~\cite{merkel2014docker}, LXC~\cite{LXC}, Singularity~\cite{kurtzer2017singularity}, Shifter~\cite{jacobsen2015contain}, and CharlieCloud~\cite{priedhorsky2017charliecloud}, 
leads us to adopt Docker, Singularity, and Shifter in the first instance.
Docker works well for local and cloud deployments, whereas
Singularity and Shifter are designed for use in HPC environments
and are supported at large-scale computing facilities (e.g., Singularity at ALCF and Shifter at NERSC). 
Singularity and Shifter implement similar models and thus 
it is easy to convert from a common representation (e.g., a Dockerfile)
to both formats. 

\name{} requires
that each container includes a base set of software, including Python~3 
and \name{} worker software.  Other system libraries or Python 
modules needed for function execution must also be included. 
When registering a function, users may optionally specify a container to be used for execution;
if no container is specified, \name{} executes functions using the worker's Python environment. 
In future work, we intend to make this process dynamic, using repo2docker~\cite{repo2docker}
to build Docker images and convert them to site-specific container formats
as needed.

\subsection{The \name{} Endpoint}\label{sec:executor}

The \name{} endpoint represents a remote resource and delivers high-performance remote execution of 
functions in a secure, scalable, and reliable manner.

The endpoint architecture, depicted in the lower portion of \figurename{~\ref{fig:arch}, 
is comprised of three components, which are discussed below:
\begin{itemize}
\item \emph{\name{} agent}: persistent process that queues and forwards tasks and results, 
interacts with resource schedulers, and batches and load balances requests.
\item \emph{Manager}: manages the resources for a single node by deploying and managing a set of workers.
\item \emph{Worker}: executes tasks within a container.
\end{itemize}
The \emph{\name{} agent} is a software agent that is deployed by a user on a 
compute resource (e.g., an HPC login node, cloud instance, or a laptop).
It registers with the \name{} service and acts 
as a conduit for routing tasks and results between the service and workers. 
The \name{} agent manages resources on its system
by working with the local scheduler or cloud API to deploy 
\emph{managers} on compute nodes.
The \name{} agent uses a pilot job model~\cite{turilli2018comprehensive} to provision 
and communicate with resources in a uniform manner,
irrespective of the resource type (cloud or cluster) or local resource manager (e.g., Slurm, PBS, Cobalt).
As each manager is launched on a compute node, it connects to and registers with the \name{} agent. 
The \name{} agent then uses ZeroMQ sockets to communicate 
with its managers. 
To minimize blocking, all communication is performed by threads
using asynchronous communication patterns. 
The \name{} agent uses a randomized scheduling algorithm to allocate tasks 
to suitable managers with available capacity.
The \name{} agent can be configured to provide 
access to specialized hardware or accelerators. 
When deploying the agent users can specify how worker
containers should be launched, enabling them to mount
specialized hardware and execute functions on that hardware. 
In future work we will
extend the agent configuration to specify custom hardware and 
software capabilities and report
this information to the \name{} agent and service for scheduling.

To provide fault tolerance and robustness, for example with respect to 
node failures, the \name{} agent relies on periodic heartbeat messages and a watchdog
process to detect lost managers. The \name{} agent tracks 
tasks that have been distributed to managers so that when failures
do occur, lost tasks can be re-executed (if permitted). 
\name{} agents communicate with the \name{} service's forwarder
via a ZeroMQ channel. 
Loss of a \name{} agent is detected by the forwarder and when 
the \name{} agent recovers, it repeats the registration 
process to acquire a new forwarder and continue receiving tasks.
To reduce overheads, the \name{} agent can shut down managers to release resources when they are not
needed; suspend managers to prevent further tasks being scheduled to them;
and monitor resource capacity to aid scaling decisions.

\emph{Managers} represent, and communicate on behalf of, the 
collective capacity of the workers on a single node, thereby limiting
the number of sockets used to just two per node. Managers determine 
the available CPU and memory resources on a node, and partition the
node among the workers. Once all workers connect to the manager it 
registers with the endpoint. Managers advertise deployed container 
types and available capacity to the endpoint. 

\emph{Workers} persist within containers and each executes one task 
at a time. Since workers have a single responsibility,
they use blocking communication to wait for functions from the manager. 
Once a task is received it is deserialized, executed,
and the serialized results are returned via the manager.

\subsection{Managing Compute Infrastructure}
\name{} is designed to support a range of computational resources, from 
embedded computers to clusters, clouds, and supercomputers,
each with distinct access modes.
As \name{} workloads are often sporadic, resources must be provisioned 
as needed to reduce costs due to idle resources.
\name{} uses \parsl{}'s provider interface~\cite{babuji19parsl} 
to interact with various resources, specify resource-specific
requirements (e.g., allocations, queues, limits, cloud instance types), 
and define rules for automatic scaling (i.e., limits and scaling aggressiveness).
This interface allows \name{} to be deployed on batch schedulers 
such as Slurm, Torque, Cobalt, SGE, and Condor; 
the major cloud vendors such as AWS, Azure, and Google Cloud; 
and Kubernetes.

\subsection{Container Management}
\name{} agents are able to execute functions on workers 
deployed in specific containers. 
Thus, managers must dynamically deploy, manage, and scale 
containers based on function requirements.  
Each manager advertises its deployed container types
to the \name{} agent. 
The \name{} agent implements a greedy, randomized scheduling algorithm to route tasks
to managers and an on-demand container deployment algorithm on the manager.
When routing functions to a manager, 
the agent attempts to send tasks to managers with suitable deployed containers. 
If there is availability on several managers, 
the agent allocates pending tasks in a randomized manner.
Upon receiving the task, the manager either
deploys a new worker in a suitable container or sends the task 
to an existing worker deployed in a suitable container. 
Both the function routing and container deployment components 
are implemented with modular interfaces via which users can 
integrate their own algorithms. 

When an endpoint is deployed on Kubernetes, both the manager and the
worker are deployed within a pod and thus the manager cannot change
worker containers. In this case, a set of managers are deployed with
specific container images and the agent simply routes
tasks to suitable managers.

\subsection{Serialization and Data Management}

\name{} supports registration of arbitrary Python functions and the 
passing of data (e.g., primitive types and complex objects) to/from functions.
\name{} uses a Facade interface that leverages several serialization libraries, including cpickle, dill, tblib, and JSON.
The \name{} serializer sorts the serialization libraries by speed and applies
them in order successively until the object is serialized.
This approach exploits the strengths of various libraries, including
support for complex objects (e.g., machine learning models) and traceback objects in a fast and transparent fashion.
Once objects are serialized, they are packed into buffers with headers that include routing tags and the serialization method,
such that only the buffers need be unpacked and deserialized at the destination.

While the serializer can act on arbitrary Python objects and input/output data, for 
performance and cost reasons we limit the size of data that can be passed through the \name{} service.
Instead, we rely on out-of-band
data transfer mechanisms, such as Globus, when passing large datasets to/from \name{} functions.
Data can be staged prior to the invocation of a function (or after the completion of a function)
and a reference to the data's location can be passed to/from the function as input/output arguments.
Many early users use this method to move large files to/from functions
(see \S\ref{sec:discussion}).

\subsection{Optimizations}

We apply several optimizations to enable high-performance function serving in a wide range of research environments. We briefly describe four optimization methods employed in \name{}.

\textbf{Container warming} is used by FaaS platforms to improve performance~\cite{wang2018peeking}. 
Function containers are kept \emph{warm} by leaving them running for a short period of time (5-10 minutes) 
following the execution of a function. 
Warm containers remove the need to instantiate a new container to execute a function, 
significantly reducing latency. 
This need is especially evident in HPC environments for several reasons: first, loading many 
concurrent Python environments and containers puts a strain on large, shared file systems; 
second, many HPC centers have their own methods for instantiating containers that may place
limitations on the number of concurrent requests; and third, individual cores are often 
slower in many core architectures like Xeon Phis. As a result the start time for containers 
can be much larger than what would be seen locally.

\textbf{Batching} requests enables \name{} to amortize costs across many function requests. 
\name{} implements two batching models:
internal batching to enable managers to request many tasks on behalf of their workers, minimizing network communication costs; 
and, a programmatic \emph{map} command that enables user-driven batching of function inputs, 
allowing users to tradeoff
efficient execution and increased per-function latency by creating fewer, larger
requests. The map command can be expressed via the SDK as: 

\noindent
\ {\small \texttt{f = fmap(func\_id, iterator, ep\_id, batch\_size, batch\_count)}},

\noindent
where \texttt{iterator} can support any Python object that implements Python's iterator interface, \texttt{batch\_size} is the number of tasks included in each batch, and \texttt{batch\_count} is the total number of batches. (Note: \texttt{batch\_count} takes precedence over \texttt{batch\_size}).
The map function 
partitions the computation's iterator into memory-efficient batches of tasks.
It exploits two key features of Python iterators:
1) iterators are evaluated in a lazy fashion and use minimal memory before being called; and
2) \emph{islice} operators can partition iterators without evaluating them.
Both batching techniques can increase overall throughput.

\textbf{Advertising with opportunistic prefetching} is a technique in which managers
continuously advertise the anticipated capacity in the near future. \name{} managers
asynchronously advertise and receive tasks, thus interleaving network communication
with computation. This can improve performance for high-throughput, short-duration, workloads.

\textbf{Memoization} involves returning a cached result
when the input document and function body have been processed previously. 
\name{} supports memoization by hashing the function body and input document and storing a mapping
from hash to computed results. Memoization is only used if explicitly set by the user.

\subsection{Security Model}
We implement a comprehensive security model to ensure 
that functions are executed by authenticated and authorized users 
and that one function cannot interfere with another. We rely on
two security-focused technologies: Globus Auth~\cite{GlobusAuth}
and containers. 

\name{} uses Globus Auth for authentication, authorization, and protection of all APIs. 
The \name{} service is registered as a \emph{resource server}, allowing users to 
authenticate using a supported Globus Auth identity (e.g., institution, Google, ORCID)
and enabling various OAuth-based authentication flows (e.g., native client)
for different scenarios. 
\name{} has associated Globus Auth scopes
(e.g., ``urn:globus:auth:scope:funcx:register\_function'')
via which other clients (e.g., applications and services) 
may obtain authorizations for programmatic access. 
\name{} endpoints are themselves Globus Auth native clients, each dependent on
the \name{} scopes, which are used to securely connect to the \name{} service. 
Endpoints require the administrator to authenticate prior to registration
in order to acquire access tokens used for constructing API requests.
The connection between the \name{}
service and endpoints is established using ZeroMQ. Communication addresses
are communicated as part of the registration process. Inbound
traffic from endpoints to the cloud-hosted service is limited to known IP addresses. 

\name{} function execution can be isolated in containers to ensure they 
cannot access data or devices outside their context.
To enable fine grained tracking of execution, 
we store execution request histories in the \name{} service
and in logs on \name{} endpoints.

\section{Evaluation}\label{sec:evaluation}

We evaluate the performance of \name{} in terms of latency, scalability, throughput, and fault tolerance. We also explore the effects of batching, memoization, and prefetching.

\subsection{Latency}

We first compare \name{} with commercial FaaS platforms by measuring the time required for single function invocations. 
We have created and deployed the same Python function on Amazon Lambda, Google Cloud Functions, Microsoft Azure Functions, and \name{}. To minimize unnecessary overhead we use the same payload when invoking each function: the string ``hello-world.'' Each function simply returns the string.

Although each provider operates its own data centers, we attempt to standardize network latencies by placing functions in an available US East region (between South Carolina and Virginia). We deploy the \name{} service and endpoint on separate AWS 
EC2 instances in the US East region.
We then measure latency as the round-trip time to submit, execute, and return a result from the function. 
We submit all requests from the login node of Argonne National Laboratory's (ANL) Cooley cluster, in Lemont, IL (18.2 ms latency to the \name{} service).

We compare the cold and warm start times for each FaaS service. 
The cold start time aims to capture the scenario where 
a function is first executed and the function code and execution environment must be
configured. To ensure cold start for \name{} functions, we restart the endpoint and measure the 
time taken to launch the first function. For the other services, we invoke functions every 15 minutes 
(providers report maximum cache times of 10 minutes, 5 minutes, and 5 minutes, for Google, Amazon, and Azure, respectively)
in order to ensure that each function starts cold.
We execute the cold start functions \num{50} times, and the warmed functions \num{10000} times.
Table~\ref{table:lat_others} shows the total time for warm and cold functions
as well as the computed overhead and function execution time. 
For the closed-source, commercial FaaS systems we obtain 
function execution time from execution logs and compute overhead
as any additional time spent invoking the function.

\begin{table}[]
\small
\caption{FaaS latency breakdown (in ms).}
\vspace{-0.1in}
\begin{tabular}{l l r r r r}
	\multicolumn{2}{l}{}    & \textbf{Overhead} & \textbf{Function} & \textbf{Total} &\textbf{Std. Dev.}\\
  \hline
	\multirow{2}*{\textbf{Azure}}      & warm &      118.0 &     12.0 &   130.0 & 14.4\\
				     & cold &    1,327.7 &     32.0 & 1,359.7  & 1,233.1 \\
	\multirow{2}*{\textbf{Google}}     & warm &      80.6 &      5.0 &   85.6  & 12.3\\
				     & cold &      203.8 &     19.0 &   222.8 & 141.8 \\
	\multirow{2}*{\textbf{Amazon}}     & warm &      100.0 &     0.3 &  100.3  & 6.9\\
				     & cold &      468.2 &     0.6 &  468.8  & 70.8\\
	\multirow{2}*{\textbf{\name{}}}    & warm &       109.1 &     2.2 &   111.3 & 11.2 \\
				     & cold &     1,491.1 & 6.1 & 1,497.2 & 10.2\\
  \hline
\end{tabular}
\label{table:lat_others}
\vspace{-0.1in}
\end{table}

Amazon Lambda, Google Functions, and Azure Functions exhibit warmed round trip times of 100ms, 86ms, and 130ms, respectively.
\name{} offers comparable performance, with 111ms round trip time.

Amazon Lambda, Google Functions, Azure Functions, and \name{} exhibit cold round trip times of 469ms, 223ms, \num{1360}ms, and \num{1497}ms, respectively. 
In the case of \name{}, the overhead is primarily due 
to the startup time of the container (see Table~\ref{table:cold_start_cost}).
Google and Amazon exhibit significantly better cold start performance than \name{},
perhaps as a result of the simplicity of our function (which requires only standard Python libraries
and therefore could be served on a standard container)
or perhaps due to the low overhead of proprietary container technologies~\cite{wang2018peeking}.

We further explore latency for \name{} by instrumenting the system.
The results in Figure~\ref{fig:lat_breakdown} for a warm
container report times as follows: 
$t_s$: Web service latency to authenticate, store the task in Redis, 
and append the task to an endpoint's queue;
$t_f$: forwarder latency to read task from Redis store, forward 
the task to an endpoint, and write the result to the Redis store; 
$t_e$: endpoint latency to receive tasks and send results to the Forwarder,  
and to send tasks and receive results from the worker; and
$t_w$: function execution time.  
We observe that $t_w$ is fast relative to the overall system latency. 
The network latency between $t_s$ and $t_f$ only includes minimal 
communication time due to internal AWS networks (measured at <1ms). 
Most \name{} overhead is captured in $t_s$ as a result
of authentication, and in $t_e$ due to internal queuing and 
dispatching. 

\begin{figure}[h]
  \includegraphics[width=0.97\columnwidth]{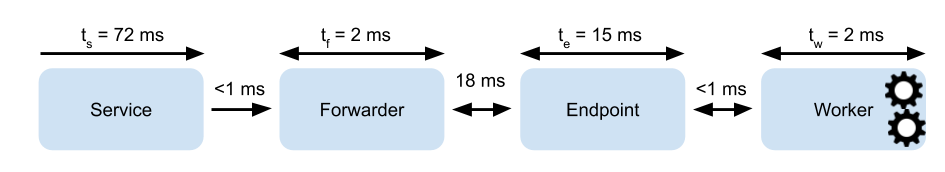}
    \vspace{-0.1in}
  \caption{\name{} latency breakdown for a warm container.}
  \label{fig:lat_breakdown}
    \vspace{-0.1in}
\end{figure}

\subsection{Scalability and Throughput}
We study the strong and weak scaling of the \name{} agent on  
ANL's Theta and NERSC's Cori supercomputers. 
Theta is a 11.69-petaflop system based on the second-generation 
Intel Xeon Phi ``Knights Landing" (KNL) processor. 
Its \num{4392} nodes each have a 64-core processor
with 16 GB MCDRAM, 192 GB of DDR4 RAM, and are interconnected with high speed InfiniBand.
Cori is a 30-petaflop system with an Intel Xeon ``Haswell" partition 
and an Intel Xeon Phi KNL partition. 
We ran our tests on the KNL partition, which 
has \num{9688} nodes, 
each with a 68-core processor (with 272 hardware threads) 
with six 16GB DIMMs, 96 GB DDR4 RAM, and interconnected with Dragonfly topology.
We perform experiments using 64 Singularity containers on each Theta node 
and 256 Shifter containers on each Cori node.
Due to a limited allocation on Cori we use the four hardware threads
per core to deploy more containers than cores.

\begin{figure}[h]
  \vspace{-0in}
  \includegraphics[width=0.9\columnwidth,trim=0.07in 0.08in 0.07in 0.08in,clip]{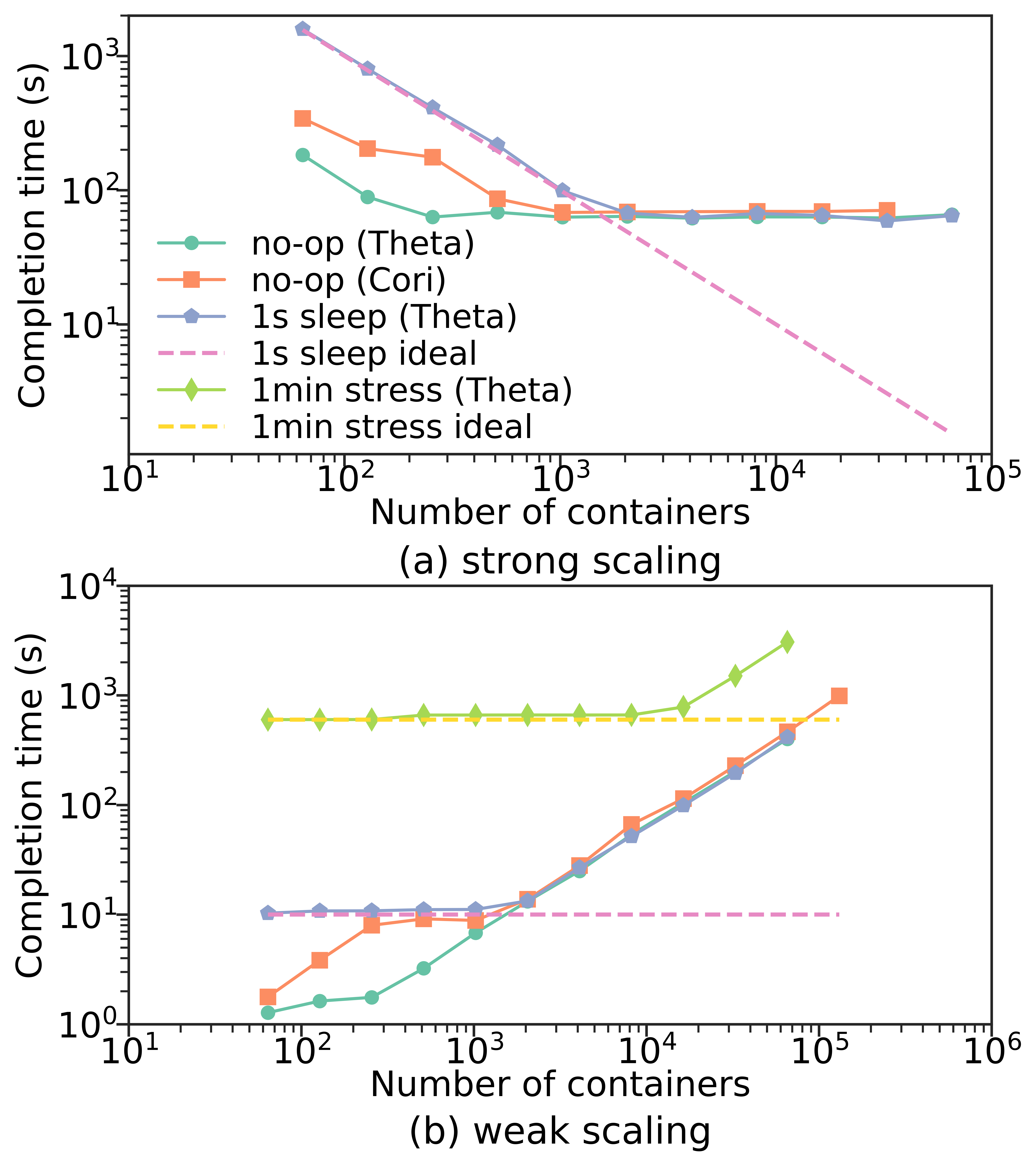}
  \vspace{-0.1in}
  \caption{Strong and weak scaling of the \name{} agent.}
\label{fig:scalability}
  \vspace{-0.1in}
\end{figure}

Strong scaling evaluates performance when the total number of function invocations is fixed; 
weak scaling evaluates performance when the average number of functions executed on each container is fixed. 
To measure scalability we created functions of various durations:  
a 0-second ``no-op'' function that exits immediately, a 1-second ``sleep'' function, and a 1-minute CPU ``stress'' function that keeps a CPU core at 100\% utilization.
For each case, we measured completion time of a batch of functions as we increased the total number of containers.
Notice that the completion time of running $M$ ``no-op'' functions on $N$ workers indicates the overhead of \name{} 
to distribute the $M$ functions to $N$ containers.
Due to limited allocation we did not execute sleep or stress 
functions on Cori, nor did we 
execute stress functions for strong scaling on Theta.

\subsubsection{Strong scaling}

Figure~\ref{fig:scalability}(a) shows the completion time of \num{100000} 
\emph{concurrent} function requests with an increasing number of containers.
On both Theta and Cori the completion time decreases as the number of containers 
increases, until we reach 256 containers for ``no-op'' and 2048 containers for ``sleep'' on Theta.
As reported by Wang et al.~\cite{wang2018peeking} and Microsoft~\cite{azureFunctionsDocs}, 
Amazon Lambda achieves good scalability for a single function to more
than 200 containers, Microsoft Azure Functions 
can scale up to 200 containers, and Google Cloud Functions does not scale well beyond 100 containers.
While these results may not indicate the maximum number of containers that can be used
for a single function, and likely include per-user limits imposed by the platform, 
our results show that \name{} scales similarly to commercial platforms.

\subsubsection{Weak scaling}
To conduct the weak scaling tests we performed \emph{concurrent} function 
requests such that each container receives, on average, 10 requests.
Figure~\ref{fig:scalability}(b) shows weak scaling for ``no-op,'' ``sleep,'' and ``stress.'' 
For ``no-op," the completion time increases with more containers on both Theta and Cori. 
This reflects the time required to distribute requests to all of the containers.
On Cori, \name{} scales to \num{131072} concurrent containers and 
executes more than 1.3 million ``no-op'' functions.
Again, we see that the completion time for ``sleep'' remains close to constant 
up to 2048 containers, and the completion time for ``stress'' remains 
close to constant up to \num{16384} containers. 
Thus, we expect a function with several
minute duration would scale well to many more containers. 

\subsubsection{Throughput}
We observe a maximum throughput for a \name{} agent (computed as number of function requests divided 
by completion time) of \num{1694} and \num{1466} requests per second on Theta and Cori, respectively.

\subsubsection{Summary}
Our results show that \name{} agents 
i) scale to \num{130000}+ containers for a single function; 
ii) exhibit good scaling performance up to approximately \num{2048} containers for a 
1-second function and \num{16384} containers for a 1-minute function; 
and iii) provide similar scalability and throughput using both Singularity and 
Shifter containers on Theta and Cori. It is important to note that these experiments study the
\name{} agent, and not the end-to-end throughput of \name{}. 
While the \name{} Web service can elastically scale
to meet demand, the communication overhead may limit throughput. To address this challenge and 
amortize communication overheads we enable batch submission of tasks. 
These optimizations are discussed in \S\ref{sec:opt}

\subsection{Elasticity}
\name{} endpoints dynamically scale and provision compute resources
in response to function load. To demonstrate this feature, we deployed a \name{} endpoint on a Kubernetes cluster, 
and used \name{} to scale the number of active pods. 
We deployed three sleep functions (running for 1s, 10s, and 20s), each in its own container. 
We limit each function to use between 0 to 10 pods. 
Every 120 seconds, we submitted one 1s, five 10s, and twenty 20s functions to the endpoint.
Figure~\ref{fig:elasticity} illustrates the concurrent functions submitted to the endpoint 
(solid lines) and the number of active pods as time elapsed (dashed lines).
We see that upon task arrivals at time 0, 120, and 240, the number of active pods is increased 
to accommodate the load. For example, at time 0, \name{} provisioned one, five, and ten (ten is the maximum) pods to process one 1s, five 10s, and twenty 20s functions, respectively.
When the functions completed, \name{} terminated unused pods.

\begin{figure}[h]
	\includegraphics[width=1\columnwidth,trim=0.05in 0.0in 0.05in 0.07in,clip]{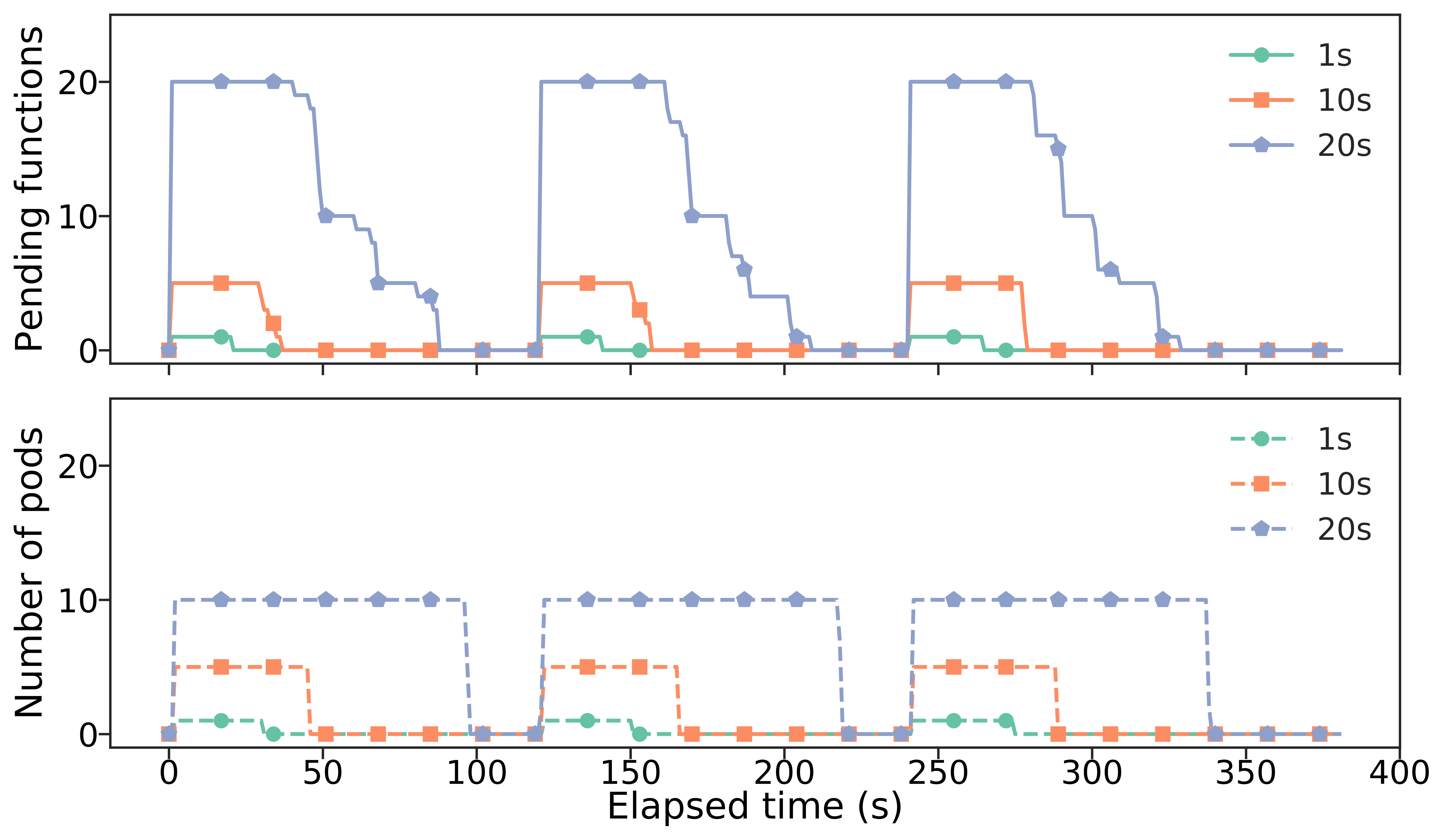}
	\vspace{-0.1in}
	\caption{Number of concurrent functions and pods over time. Top: number of pending and executing functions. Bottom: number of active pods serving functions.}
	\label{fig:elasticity}
	\vspace{-0.1in}
\end{figure}

\subsection{Fault Tolerance}
\name{} uses heartbeat messages to detect and respond to component failures.
We evaluate this feature by forcing endpoint and manager failures while processing
a workload of 100ms sleep functions launched at a uniform rate.

The first experiment uses two managers processing a stream of tasks launched at uniform intervals, ensuring that the system is kept at capacity. One manager is terminated after 2 seconds and restarted after 4 seconds. Figure~\ref{fig:fault_tolerance} illustrates the task latencies measured as the experiment progresses. It shows that task latency increases immediately following the failure, as tasks are queued, and then quickly reduce after the manager recovers.

To explore the impact of an endpoint failing (or going offline), we launch a stream of tasks at a uniform rate, and trigger the failure and recovery of the endpoint after 43s and 85s, respectively.
Figure~\ref{fig:ep_fault_tolerance} illustrates the task latencies measured as the experiment progresses. We see that task latency increases immediately following the failure and returns to previous levels after recovery.

\begin{figure}[h]
  \includegraphics[width=0.8\columnwidth,trim=0.1in 0.1in 0.08in 0.07in,clip]{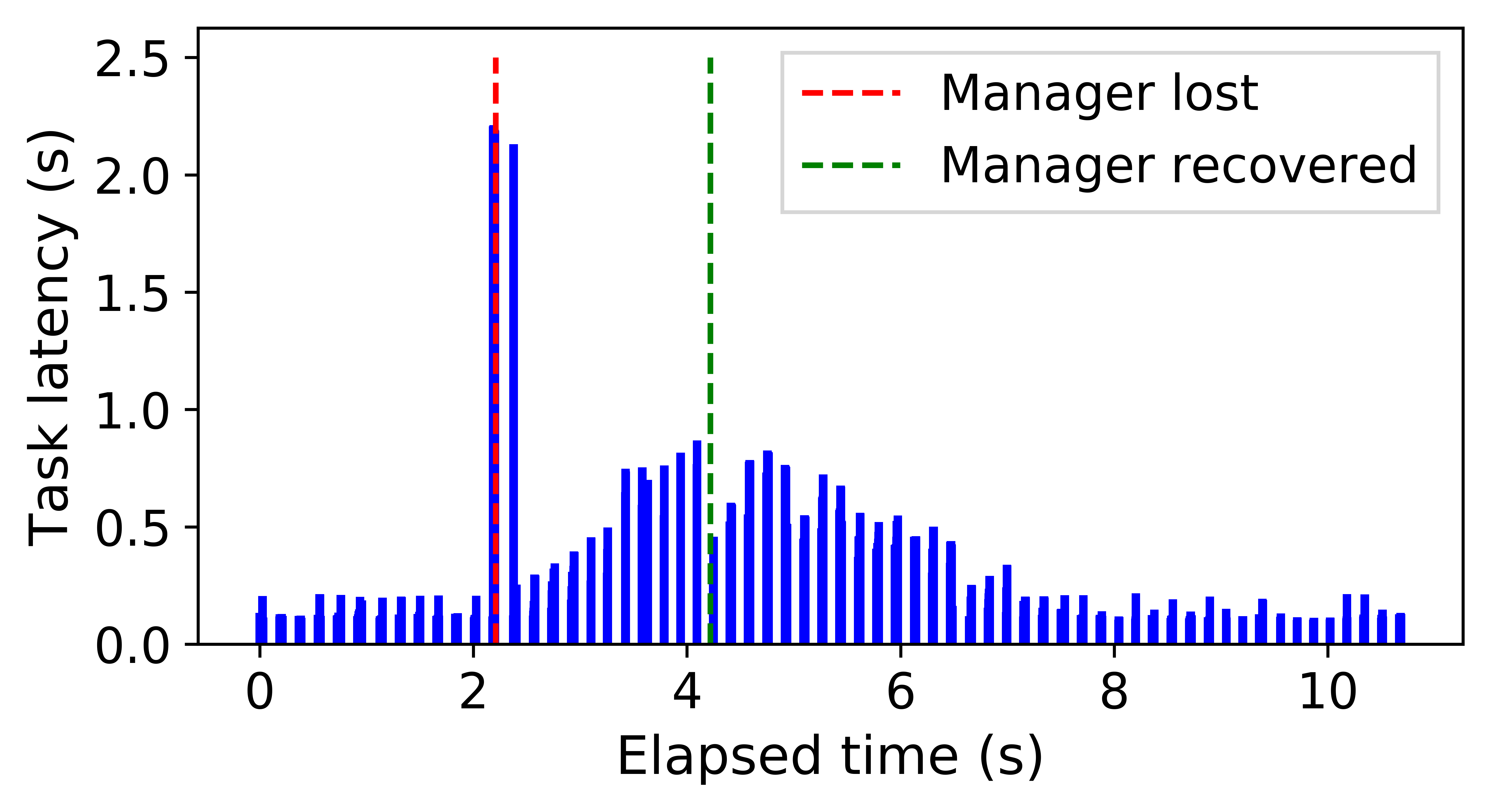}
    \vspace{-0.1in}
  \caption{Timeline showing task processing latency for 100ms functions, when a manager fails and recovers.}
\label{fig:fault_tolerance}
  \vspace{-0.1in}
\end{figure}

\begin{figure}[h]
  \includegraphics[width=0.8\columnwidth,trim=0.1in 0.1in 0.08in 0.07in,clip]{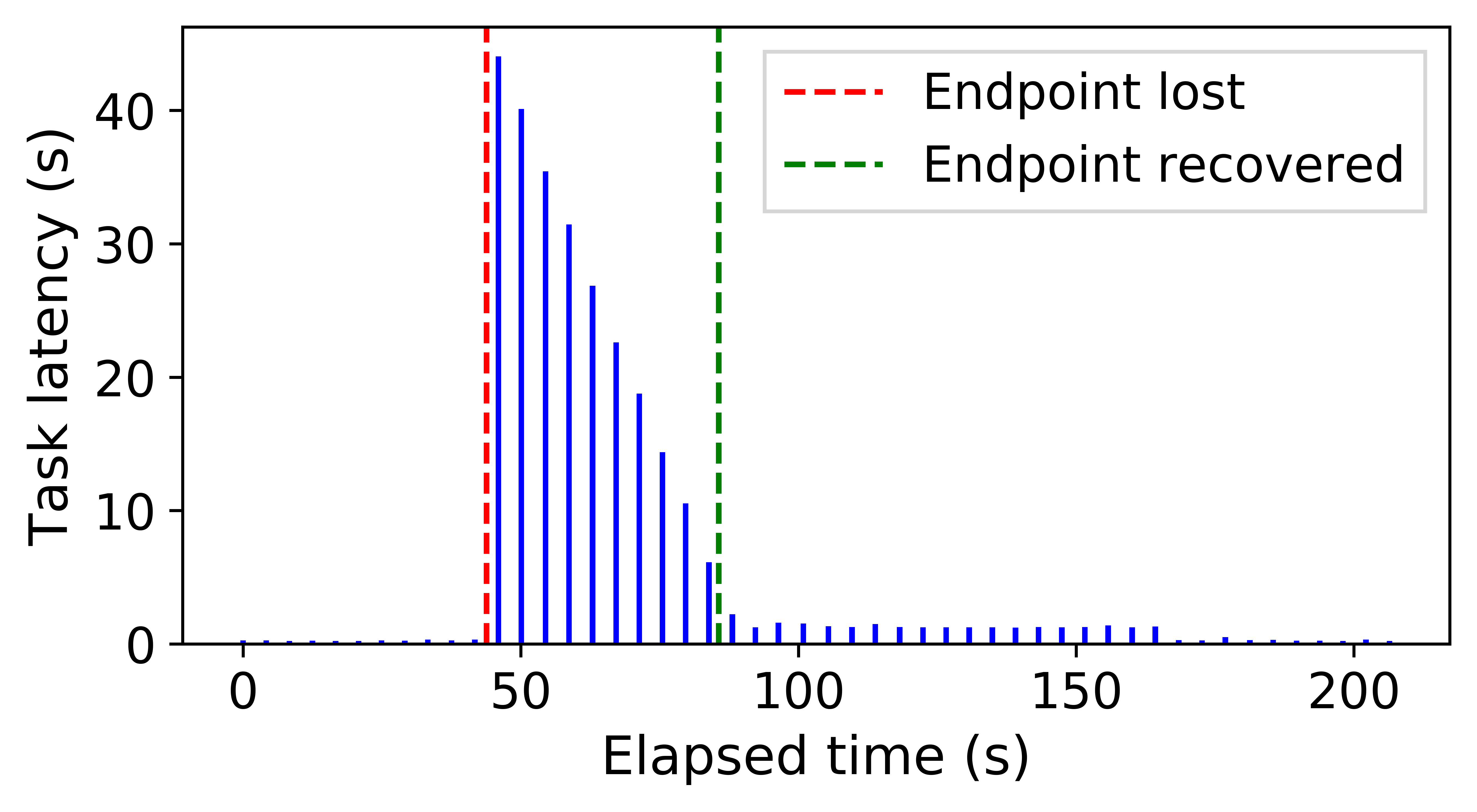}
    \vspace{-0.1in}
  \caption{Timeline showing task processing latency for 100ms functions, when an endpoint fails and recovers.}
\label{fig:ep_fault_tolerance}
  \vspace{-0.1in}
\end{figure}

\subsection{Optimizations}
\label{sec:opt}
In this section we evaluate the effect of our optimization mechanisms. In particular, we investigate how container initialization, batching, prefetching, and memoization impact performance.

\subsubsection{Container instantiation}

To understand the time to instantiate various container technologies on different execution resources
we measure the time it takes to start a container and execute a Python command that imports \name{}'s worker modules---the
baseline steps that would be taken by every cold \name{} function.
We deploy the containers on an AWS EC2 \texttt{m5.large} instance and on compute nodes on Theta and Cori following best practices
laid out in facility documentation. \tablename~\ref{table:cold_start_cost} shows the results.
We speculate that the significant performance deterioration of container instantiation on HPC systems can be attributed
to a combination of slower clock speed on KNL nodes and shared file system contention when fetching images.
These results highlight the need to apply function warming approaches to reduce overheads.

\begin{table}[h]
\caption{Cold container instantiation time for different container technologies on different resources.}
\label{extractor-tab}
 \vspace{-0.1in}
\begin{center}
  \begin{tabular}{l l c c c}
    \textbf{System} & \textbf{Container} & \textbf{Min (s)} & \textbf{Max (s)} & \textbf{Mean (s)} \\
    \hline
    Theta & Singularity & 9.83 & 14.06 & 10.40  \\
    Cori & Shifter & 7.25     & 31.26    & 8.49 \\
    EC2 & Docker   & 1.74     & 1.88     & 1.79 \\
    EC2 & Singularity & 1.19  & 1.26     & 1.22 \\
    \hline
  \end{tabular}
\end{center}
  \vspace{-0.1in}
\label{table:cold_start_cost}
\end{table}

\subsubsection{Executor-side batching}
To evaluate the effect of executor-side batching we submit \num{10000} concurrent ``no-op'' function requests
and measure the completion time when executors can request one function at a time (batching disabled) vs
when they can request many functions at a time based on the number of idle containers (batching enabled). 
We use 4 nodes (64 containers each) on Theta.
We observe that the completion time with batching enabled is 6.7s (compared to 118s when disabled).

\subsubsection{User-driven batching}
Figure~\ref{fig:map} shows the strong-scaling performance of \name{}'s map command as we vary batch
size and number of workers. 
In this experiment we launch 10 million functions each executing for 10$\mu$s,
with the client and endpoint both running on one AWS EC2 c5n.9xlarge instance.
We see that \name{} can achieve a peak throughput of 1.2 million 
functions-per-second on a single machine, 
well beyond what is possible without batching.

\begin{figure}[h]
  \vspace{-0.1in}
  \includegraphics[width=0.9\columnwidth]{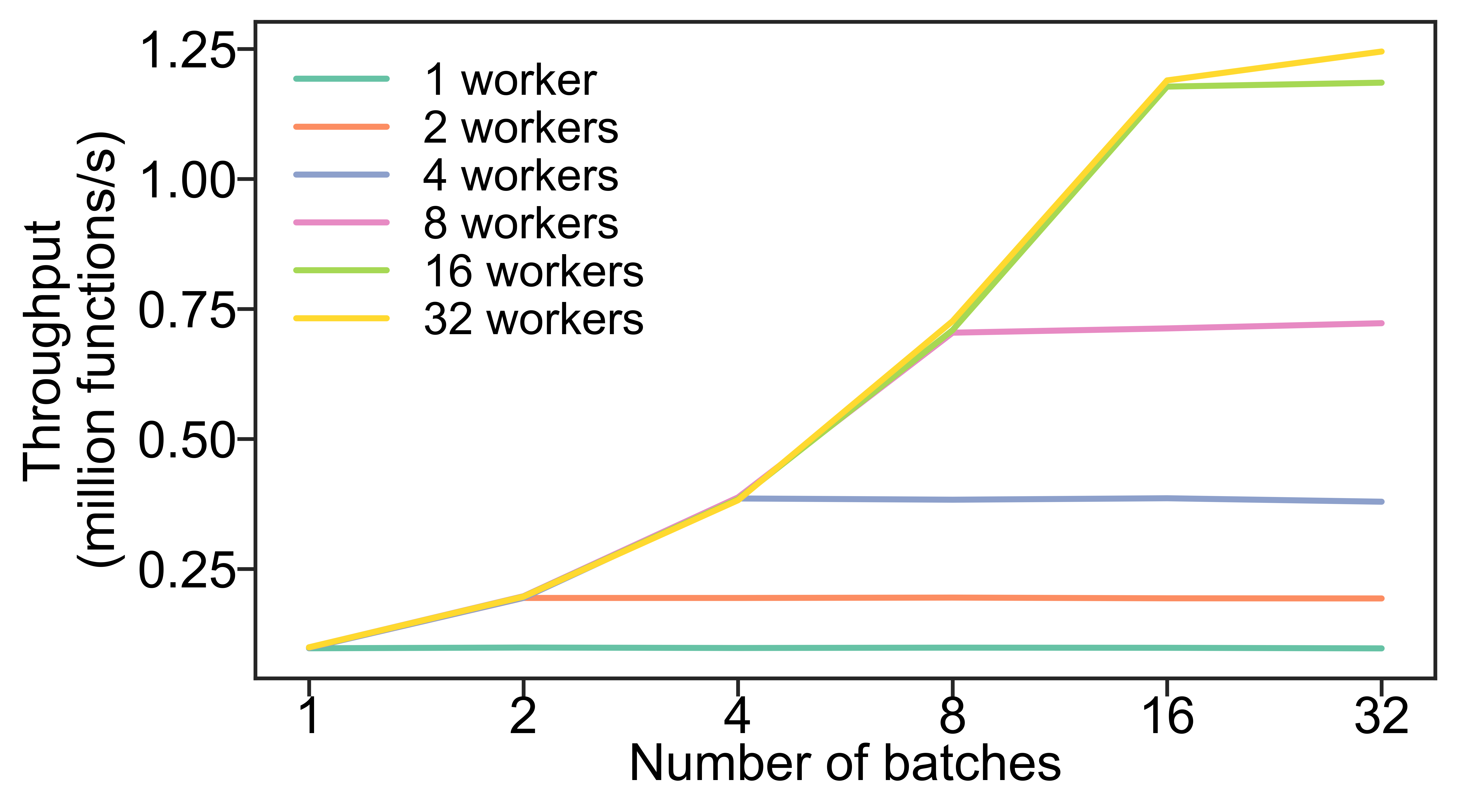}
  \vspace{-0.1in}
  \caption{Strong scaling performance over 10M functions} 
  \label{fig:map}
\end{figure}

\subsubsection{Batching case studies}
To evaluate the effect of user-driven batching we explore a subset of the scientific case studies 
discussed in \S\ref{sec:requirements}. These case studies
represent various scientific functions, ranging in execution time from half a second through to almost one minute, and provide perspective to the real-world effects of batching on different types of functions.
The batch size is defined as the number of requests transmitted to the container for execution. 
Figure~\ref{fig:batching} shows the average latency per request (total completion time of the 
batch divided by the batch size), as the batch size increases. We observe that batching provides 
enormous benefit for the shortest running functions and reduces the average latency dramatically when 
combining tens or hundreds of requests. However, larger batches provide little benefit,
indicating that it would be better to distribute the requests to additional workers. 
Similarly, long-running functions do not benefit, as the communication and startup costs are small relative to computation time.

\begin{figure}[h]
	  \vspace{-0.1in}
	\includegraphics[width=0.8\columnwidth]{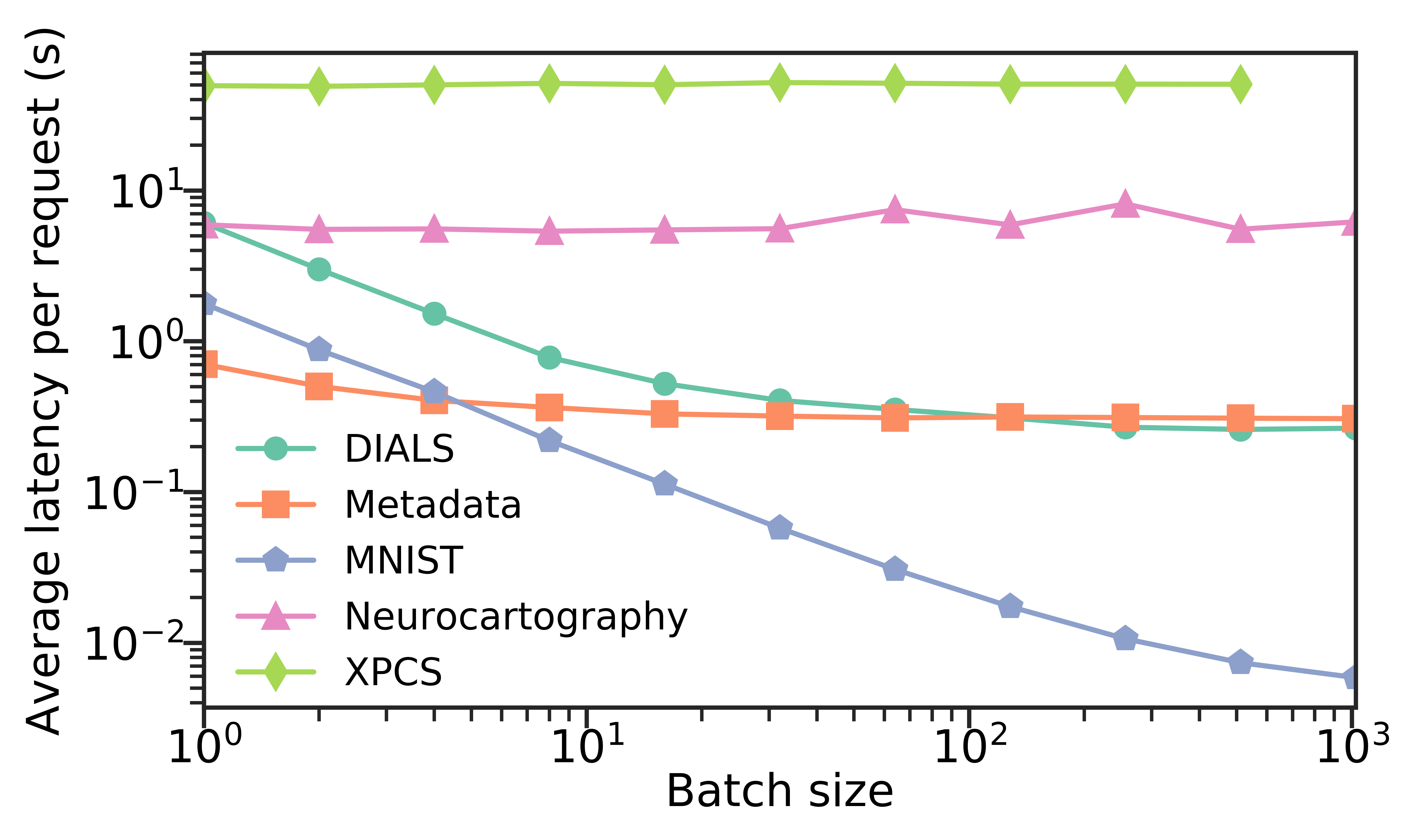}
		  \vspace{-0.1in}
	\caption{Effect of batch size (1--1024) on the use cases.}
	\label{fig:batching}
	  \vspace{-0.1in}
\end{figure}

\subsubsection{Prefetching}
We evaluate the effect of prefetching by creating a no-op and 1ms, 10ms, and 100ms sleep functions, and measuring the time for \num{10000} concurrent function requests as the prefetch count per node is increased.
Figure~\ref{fig:prefetching} shows the results of each function with 4 nodes (64 containers each) on Theta. 
We observe that completion time decreases dramatically as prefetch count increases. 
This benefit starts diminishing when prefetch count is greater than 64, suggesting that a good setting of prefetch count would be close to the number of containers per node. 
\begin{figure}[h]
	\vspace{-0.1in}
	\includegraphics[width=0.8\columnwidth]{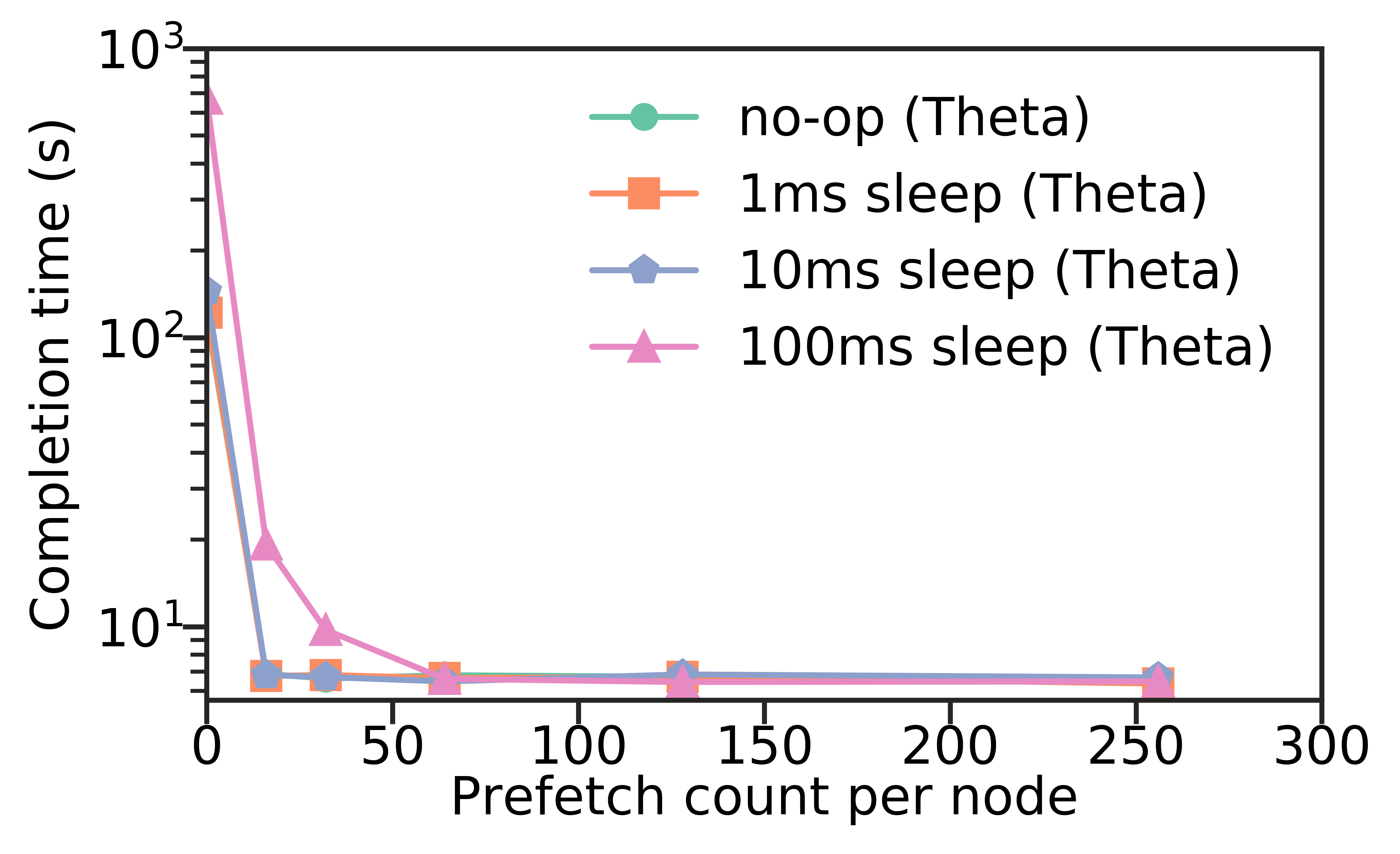}
	\vspace{-0.1in}
	\caption{Effect of prefetching.}
	\label{fig:prefetching}
	\vspace{-0.1in}
\end{figure}

\subsubsection{Memoization}
To measure the effect of memoization, we create a function that sleeps for one second 
and returns the input multiplied by two. We submit \num{100000} concurrent function requests to \name{}.
Table~\ref{table:memoization} shows the completion time of the \num{100000} requests 
when the percentage of repeated requests is increased. 
We see that as the percentage of repeated requests increases, the completion time 
decreases dramatically. This highlights the significant performance benefits of memoization
for workloads with repeated deterministic function invocations.
\begin{table}[h]
	\caption{Completion time vs.\ number of repeated requests.}
	\vspace{-0.1in}
	\begin{center}
		\begin{tabular}{c c c c c c}
			\hline
			\textbf{\begin{tabular}[c]{@{}l@{}}Repeated requests (\%)\end{tabular}}
				 & 0 & 25 & 50 & 75 & 100 \\
			\hline
			\textbf{Completion time (s)} & 403.8&  318.5  &    233.6 &  147.9 &   63.2 \\
			\hline
		\end{tabular}
	\end{center}
	\vspace{-0.1in}
	\label{table:memoization}
\end{table}

\section{Experiences with \name{}}
\label{sec:discussion}

We conclude by describing our experiences applying \name{} to the six
scientific case studies presented in \S\ref{sec:requirements}. 

\textbf{Metadata extraction:}
\xtract{} uses \name{} to execute its pre-registered metadata extraction 
functions centrally (by transferring data to the service)
and on remote \name{} endpoints where data reside without moving them to the cloud.  

\textbf{Machine learning inference:}
\dlhub{} uses \name{} to perform model inference on arbitrary compute resources. 
Each model is registered as a \name{} function, mapped to the \dlhub{} registered containers. 
\name{} provides several advantages to \dlhub{}, most notably, that it allows
\dlhub{} to use various remote compute resources via a simple interface, 
and includes performance optimizations (e.g., batching and caching) that improve overall inference performance.

\textbf{Synchrotron Serial Crystallography (SSX):} 
We deployed the DIALS~\cite{waterman2013dials} crystallography processing tools as 
\name{} functions. \name{} allows SSX researchers to submit the same 
\emph{stills process} function to either a local endpoint to perform data 
validation or HPC resources 
to process entire datasets and derive crystal structures.

\textbf{Quantitative Neurocartography:} 
Previous practice
depended on batch computing jobs that 
required frequent manual intervention for authentication, configuration, and failure resolution. 
With \name{}, researchers can
use a range of computing resources without the overheads previously associated with
manual management. In addition, they can now
integrate computing into their automated visualization and analysis workflows 
via programmatic APIs. 

\textbf{X-ray Photon Correlation Spectroscopy (XPCS):} 
We incorporated the XPCS-eigen corr function, deployed as a \name{} function, into an on-demand analysis pipeline triggered as data are collected at the beamline.
This work allows scientists to offload analysis tasks to HPC resources, simplify large-scale
parallel processing for large data rates. \name{}'s scalability meets the demands of the XPCS data rates by
acquiring multiple nodes to serve functions.

\textbf{Real-time data analysis in High Energy Physics (HEP)}:
We developed a \name{} backend 
to Coffea~\cite{coffea}, an analysis framework that can be used to parallelize real-world HEP analyses operating on columnar data to aggregate histograms of analysis products of interest in real time. Subtasks representing partial histograms are dispatched as \name{} requests. We completed a typical HEP analysis of 300 million events in nine minutes (1.9 $\mu$s/event), simultaneously using two \name{} endpoints provisioning heterogeneous resources.

\textbf{Summary:}
Based on discussion with these researchers we have identified several 
benefits of the \name{} approach in these scenarios. 
1) \name{} abstracts the complexity of using diverse compute resources. 
Researchers are able to incorporate scalable analyses
without having to know anything about the computing environment 
(batch queues, container technology, etc.). 
2) Researchers appreciated the ability to simplify application code, 
automatically scale resources to workload needs, and avoid the complexity
of mapping applications to batch jobs. Several highlighted the benefits
for elastically scaling resources to long-tail task durations. 
3) Researchers
found that the flexible web-based authentication model significantly
simplified remote computing when compared to the previous models
that relied on SSH keys and two-factor authentication.
4) Several case studies use \name{} to enable event-based processing. 
We found that the \name{} model lends itself well to such use cases, 
as it allows for the execution of sporadic workloads. 
The neurocartography, XPCS, and SSX use cases all exhibit such characteristics, 
requiring compute resources only when experiments are running. 
5) Researchers highlighted portability as a benefit of \name{},
not only for using multiple resources but also to overcome scheduled
and unscheduled facility downtime. 
6) \name{} allowed resources to be used efficiently and opportunistically, for example
using backfill queues to quickly execute tasks.
7) \name{} allowed users to securely share their functions, enabling 
collaborators to easily (without needing to setup environments) apply
functions on their own datasets. This was
particularly useful in the XPCS use case as researchers share
access to the same instrument.

While initial feedback has been encouraging, our experiences also highlight
important challenges that need to be addressed. 
1) FaaS is not suitable for some applications, for 
example applications with tightly integrated computations, that share large amounts of data,
and are implemented with large and complex code bases.
2) Containerization does not always provide entirely portable 
codes that can be run on arbitrary resources, due to the need to compile 
and link resource-specific modules. For example, in the XPCS use case
we needed to compile codes specifically for a target resource. 
3) The coarse allocation models employed by research infrastructure
does not map well to fine grain and short duration function usage, 
work is needed to support accounting and billing models to track 
usage on a per-user and per-function basis.
4) There are other barriers that make it difficult to decompose applications into functions. 
For example, the neurocartography tools are designed to perform many interlaced tasks
and thus we chose to  
package the entire toolkit as a function rather than to decompose these tools
into many functions. 
We also found that it can be difficult to modify applications for 
stateless execution, as state is not easily shared between
functions, and poorly designed solutions may lead to significant communication overhead.

\section{Related Work}\label{sec:survey}

FaaS platforms have proved extremely successful in industry as a way
to reduce costs and remove the need to manage infrastructure.

\textbf{Hosted FaaS platforms:}
\emph{Amazon Lambda}~\cite{amazonlambda},
\emph{Google Cloud Functions}~\cite{googlecloudfunctions}, and \emph{Azure Functions}~\cite{azureFunctions} are the most well-known FaaS platforms. Each service supports various function languages
and trigger sources, connects directly to other cloud services, and is billed in granular increments.  Lambda uses Firecracker, a custom virtualization technology built on KVM, to create lightweight micro-virtual machines. 
To meet the needs of IoT use cases,
some cloud-hosted platforms also support local deployment (e.g., AWS Greengrass~\cite{greengrass}); 
however, they support only single machines and require that functions be exported from the cloud platform.

\textbf{Open source platforms:}
Open FaaS platforms resolve two of the key challenges to using FaaS for scientific workloads: they 
can be deployed on-premise and can be customized to meet the requirements of data-intensive 
workloads without set pricing models.

\emph{Apache OpenWhisk}~\cite{openwhisk}, the basis of IBM Cloud Functions~\cite{IBMCloudFunctions},
defines an event-based programming model, consisting of \emph{Actions} which are stateless, runnable functions, \emph{Triggers} which are the types of events OpenWhisk may track, and \emph{Rules} which associate one trigger with one action. OpenWhisk can be deployed locally as a service using a Kubernetes cluster. 

\emph{Fn}~\cite{Fn} is an event-driven FaaS system that executes functions in Docker containers. Fn allows users to logically group functions into applications. 
Fn can be deployed locally (on Windows, MacOS, or Linux) or on Kubernetes.

The \emph{Kubeless}~\cite{Kubeless} FaaS platform builds upon Kubernetes. 
It uses Apache Kafka for messaging, provides a CLI that mirrors Amazon Lambda, and supports comprehensive monitoring. 
Like Fn, Kubeless allows users to define function groups that share resources.

\emph{SAND}~\cite{akkus2018sand} is a lightweight, low-latency FaaS platform from Nokia Labs that provides application-level sandboxing and a hierarchical message bus. 
SAND provides support for function chaining via user-submitted workflows. 
SAND is closed source and as far as we know cannot be downloaded and installed locally.

\emph{Abaco}~\cite{stubbs2017containers} implements the Actor model, where an \textit{actor} is an Abaco runtime mapped to a specific Docker image. Each actor executes in response to messages posted to its \textit{inbox}. It supports functions written in several programming languages and automatic scaling. Abaco also provides fine-grained monitoring of container, state, and execution events and statistics. Abaco is deployable via Docker Compose. 

\textbf{Comparison with \name{}:}
Hosted cloud providers implement high performance and reliable FaaS models 
that are used by an enormous number of users. However, they are not designed
to support heterogeneous resources or research CI (e.g., schedulers, containers), 
do not integrate with the science ecosystem 
(e.g., in terms of data and authentication models), and can be costly. 

Open source and academic frameworks support on-premise deployments and can be 
configured to address a range of use cases. However, each system we surveyed
is Docker-based 
and relies on Kubernetes (or other
container orchestration platforms) for deployment. 
These systems therefore cannot be easily adapted to existing HPC environments. 
We are not aware of any systems that support remote execution
of functions over a distributed or federated ecosystem of endpoints.

\textbf{Other Related Approaches}
FaaS has many predecessors, notably
grid and cloud computing, container orchestration, and analysis systems. 
Grid computing~\cite{Foster2001} laid the foundation for 
remote, federated computations, most often through
federated batch submission~\cite{krauter02gridmanagement}. 
GridRPC~\cite{seymour02gridrpc} defines an API for executing 
functions on remote servers requiring that developers implement
the client and the server code. 
\name{} extends these ideas to allow interpreted functions
to be registered and subsequently dynamically 
executed within sandboxed containers via a standard endpoint API.

Container orchestration systems~\cite{rodriguez19containers, hightower17kubernetes, hindman11mesos} allow
users to scale deployment of containers while 
managing scheduling, fault tolerance, resource provisioning, and addressing
other user requirements. 
These systems primarily rely on dedicated, cloud-like
infrastructure and cannot be directly used with most HPC resources. 
However, these systems provide a basis for other serverless platforms, such as Kubeless. 
\name{} focuses at the level of scheduling and managing functions, that are deployed
across a pool of containers. 
We apply approaches from container orchestration
systems (e.g., warming) to improve performance.

Data-parallel systems such as Hadoop~\cite{hadoop} and Spark~\cite{spark}
enable map-reduce style analyses. 
Unlike \name{}, these systems dictate a particular programming
model on dedicated clusters. 
Python parallel computing libraries such as Parsl and Dask~\cite{dask} 
support development of parallel programs, 
and parallel execution of selected functions within those scripts, 
on clusters and clouds. These systems could be extended to use \name{}
for remote execution of tasks.

\section{Conclusion}\label{sec:conclusion}

\name{} is a distributed FaaS platform that is designed to support the unique
needs of scientific computing. It combines a reliable and easy-to-use cloud-hosted
interface with the ability to securely execute functions on distributed
endpoints deployed on various computing resources. 
\name{} supports many HPC systems and cloud platforms, can use three container 
technologies, and can expose access to heterogeneous and specialized computing resources.  
We demonstrated that \name{} provides comparable latency 
to that of cloud-hosted FaaS platforms and showed that \name{} agents can 
scale to execute 1M tasks over \num{130000} concurrent workers 
when deployed on the Cori supercomputer.  
We also showed that \name{} can elastically scale in response to 
load, automatically respond to failures, 
and that user-driven batching 
can execute more than one million functions per second on a single machine.

\name{} demonstrates the advantages of adapting the FaaS model to create
a federated computing ecosystem. Based on early experiences using \name{} in 
six scientific case studies, we have found that the approach
provides several advantages, 
including abstraction, code simplification, portability, scalability, and sharing; however, 
we also identified several limitations including suitability for some applications, 
conflict with current allocation models, and challenges decomposing
applications into functions. 
We hope that \name{} will serve as a flexible platform for scientific
computing while also enabling new research related to function scheduling, 
dynamic container management, and data management.
 
In future work we will extend \name{}'s container management capabilities to 
create containers dynamically based on function requirements,
and to stage containers to endpoints on-demand.
We will also explore techniques for optimizing performance, for example by
sharing containers among functions with similar dependencies and developing
resource-aware scheduling algorithms. 
\name{} is open source and available at \url{https://github.com/funcx-faas}.

\section*{Acknowledgment}
This work was supported in part by Laboratory Directed Research and
Development funding from Argonne National Laboratory under U.S. Department of
Energy under Contract DE-AC02-06CH11357 and used resources of the
Argonne Leadership Computing Facility.

\bibliographystyle{ACM-Reference-Format}
\bibliography{main}
\end{document}